
\input phyzzx

\def\prl{Phys. Rev. Lett. }

\overfullrule=0pt
\tolerance=5000
\overfullrule=0pt
\twelvepoint

\REF\GP {{\it The Quantum Hall Effect},
Springer-Verlag, New York, Heidelberg,
1990. edited by R. Prange and S. Girvin.  2nd ed.}
\REF\ZHK {S.C. Zhang, T.H. Hansson, and S. Kivelson, Phys. Rev.
Lett. {\bf 62}(1989)82.}
\REF\RE {N. Read, Phys. Rev. Lett. {\bf 62}(1989)86.}
\REF\BW {B. Blok, X.G. Wen, Phys. Rev. {\bf B42}(1990)8145; ibid. {\bf
43}(1991)8337.}
\REF\REA { N. Read, Phys. Rev. Lett. {\bf 65}(1990)1502.}
\REF\WQ{X.G. Wen and Q. Niu, Phys. Rev. {\bf B41}(1990)9377.}
\REF\GREAD{G. Moore and N. Read, Nucl. Phys. {\bf 360B}(1991)362.}
\REF\WI{{\it Fractional Statistics and  Anyon Superconductivity},
World Scientific, 1990. edited by F. Wilczek.}
\REF\ROBERTO{R. Iengo and K. Lechner, Phys. Rep. {\bf 213}, 179 (1992).}
\REF\LAO{R.B. Laughlin, \prl {\bf 50}(1983)1395.}
\REF\HALD{F.D.M. Haldane, \prl {\bf 51}(1983)605.}
\REF\HR{ F.D.M. Haldane and E.H. Rezayi, Phys. Rev. {\bf B31} (85) 2529.}
\REF\HALP{B.I. Halperin, Phys. Rev. Lett. {\bf 52}(1984)1583.}
\REF\LIN{Dingping Li, {\it Anyons and Quantum Hall Effect on the Sphere},
SISSA/ISAS/129/91/EP.}
\REF\LINN{Dingping Li, Phys. Lett. {\bf A169}(1992)82.}
\REF\MACO{Marco A.C. Kneipp, Int. J. Mod. Phys. {\bf A7}(1992)4477.}
\REF\CONWAY{J.H. Conway and N.J.A. Sloane, {\it Sphere Packings, Lattices and
Groups}. Springer-verlag, 1988.}
\REF\HRN{ F.D.M. Haldane, Phys. Rev. Lett {\bf 55} (85) 2095.}
\REF\ILN {R. Iengo and K. Lechner, Nucl. Phys. {\bf B365}(1991)551.}
\REF\PRE{Dingping Li, {\it Hierarchical Wave Functions and Fractional
Statistics in Fractional Quantum Hall Effect on the Torus},
SISSA/ISAS/58/92/EP.}
\REF\FRACT{R.B. Laughlin, ``Fractional Statistics in the Quantum Hall Effect",
in [\WI].}
\REF\XIE{F.C. Zhang and X.C. Xie, Phys. Rev. {bf B40}(1989)11449.}
\REF\NAP{G. Cristofano, G. Maiella, R. Musto and F. Nicodemi, {\it Hall
Conductance and Edge States in the Coulomb Gas Vertex Operator Formalism}.
DSF-T-91/6 and INFN-NA-IV-91/6.}

\pubnum{SISSA/ISAS/2/92/EP, cond-mat/9212034}
\date{January, 1992}
\titlepage
\title{Hierarchical Wave Functions of Fractional Quantum
Hall Effect on the Torus}
\vglue-.25in
\author{Dingping Li\foot{emai: LIDP@TSMI19.SISSA.IT}}
\address{ International School for Advanced Studies,
I-34014 Trieste, Italy}
\bigskip
\abstract{One kind of the hierarchical wave functions of Fractional Quantum
Hall Effect on the torus is constructed.
We find that the wave functions  closely relate
to the wave functions of generalized Abelian Chern-Simons theory.}
\endpage

The past ten years have seen a great deal of interest in Fractional Quantum
Hall Effect not only in the group of the condense matter physicists, but
also in the group of the mathematical physicists.
In especial the relation between FQHE and Chern-Simons-Conformal Field Theory
(Chern-Simons-Conformal Field Theory plays an important role in the recent
developments of the  theoretical
physics), was uncovered in the past few years
[\GP,\ZHK,\RE,\BW,\REA,\WQ,\GREAD,etc.].
FQHE is a place which can realize some theoretical
ideas in the theoretical physics
on nature. For example, the fractionally charged quasiparticle excitations
obey fractional statistics. Such kinds of the excitations are called anyons,
which were discovered theoretically by J.M. Leinaas and J. Myrheim and made
popular by  F. Wilczek ([\WI] and references in it; see also [\ROBERTO]).
In this paper, we shall investigate
the relation between FQHE and Chern-Simons theory.
\par
The Laughlin wave functions [\LAO] of
FQHE on compact surfaces, for example, sphere and torus
at the filling ${1\over m}$ with $m$ being an odd
integer has been obtained in [\HALD] by F.D.M. Haldane and in [\HR] by
F.D.M. Haldane and E.H.  Rezayi.
The hierarchical wave functions of FQHE have been discussed in
[\HALD,\HALP].
The hierarchical wave functions of FQHE have been investigated most
recently in [\BW,\REA]
(on the sphere, those wave function can be obtained  by following  the method
developed in [\LIN,\LINN]).
Then we may ask whether it is possible to extend the
result in [\HR] to get the hierarchical wave functions on the torus.
\par
The results of this paper are as follows.
(i) the result in [\HR] is reviewed.
The ground wave functions are reformulated  in  a new bases.
Then we proceed to get the hierarchical wave functions on the torus
(this kind of  wave function on the plane has been studied in [\REA]).
The degeneracy of the wave functions
is obtained and the result agrees with the prediction in
[\WQ,\GREAD].   (ii)The wave functions are found to be
closely related with the wave functions of the generalized Abelian
Chern-Simons theory [\MACO].
\par
Following [\HR], we consider a magnetic field with potential
${\bf A}=- B y $\^x, the wave function describing a electron in the
lowest Landau level has the form
$$\psi (x,y)=e^{-{ B y^2\over 2}}f(z) \, , \eqn\aaa$$
where $f(z)$ is the holomorphic function, and the units $e=1 \, , \hbar =1 $
are used. The lagrangian of the electron in the magnetic field is
$$L=\sum_{i=1,2}{1\over 2}m {(v^i)}^2 +A^iv^i \, , \eqn\aab$$
where $L$ is invariant up to a total time derivative under the
translations. The corresponding Noether currents due to the translations are
$$t_x=m{\dot x}-By \, , t_y=m{\dot y}+Bx \, . \eqn\aac$$
The conjugate momenta are
$$p_x=m{\dot x}-By \,, p_y=m{\dot y} \, .\eqn\aad$$
So
$$t_x=p_x \, , t_y=p_y+Bx \, . \eqn\aae$$
They commute with Hamiltonian
$$H={1\over 2m} [{(p_x+By)}^2+{(p_y)}^2] \, ,\eqn\aaf$$
with the commutations $[x\, ,p_x]=i\, ,[y\, ,p_y]=i$ when the theory is
quantized.  By identifying
$z \sim z+m+n \tau$ with $\tau =\tau_1 +i\tau_2$ and $\tau_2 \geq 0$,
we obtain a torus specified by $\tau$.
The consistent boundary conditions imposed on the wave function of
the electron on this torus are
$$e^{it_x} \psi = e^{i\phi_1} \psi \, , e^{i\tau_1 t_x +i\tau_2 t_y}
\psi = e^{i\phi_2} \psi \, , \eqn\aag$$
with the condition $\tau_2 B=2\pi \Phi$, and $\Phi$
is an integer, which will insure that $e^{it_x}, e^{i\tau_1 t_x +i\tau_2 t_y}$
commute with each other for the consistence of the equation \aag.
By using the relation
$$e^{i\tau_1 t_x +i\tau_2 t_y}=e^{-i\tau_2 Bx^2 \over 2\tau_1}
e^{i\tau_1 p_x +i\tau_2 p_y}e^{i\tau_2 Bx^2 \over 2\tau_1}\, ,$$
\aag can be written as
$$f(z+1)=e^{i\phi_1} f(z) \, ,
f(z+\tau )=e^{i\phi_2}e^{-i\pi \Phi (2z+\tau)}f(z) \, . \eqn\aah$$
For the many-particle wave functions, the condition of the equation \aah
is imposed on every particle. Our task is to seek Laughlin-Jastrow
wave functions for FQHE.
\par
Now let us give a brief review about the definition of  $\theta$ function.
The standard $\theta$ function is defined as
$$\theta (z|\tau ) =\sum _n \exp (\pi in^2 \tau +2\pi inz)\, ,
n\subset integer \, . \eqn\aai$$
More generally, $\theta$ function on the lattice [\CONWAY] is given by
$$\theta (z|e, \tau ) =\sum _{v} \exp (\pi i v^2 \tau +2\pi iv\cdot z)
\, , \eqn\aaj$$
where $v$ is a vector on a {\it l-dimension} lattice, $v=\sum_{i=1}^l n_i e_i$,
with $n_i$ being integers,  $e_i \cdot e_j =\Lambda_{ij}$ and $z=z_ie_i$.
The $\theta$ function in the equation \aai is
the $\theta$ function defined by \aaj with
$l=1, e_1 \cdot e_1 =1$.    Furthermore we define
$$\theta {a\brack b} (z|e, \tau ) =\sum _{v}
\exp (\pi i {(v+a)}^2 \tau +2\pi i(v+a)\cdot (z+b))\, , \eqn\aak$$
where $a, b$ are  arbitrary vectors on the lattice. Only the
positive lattice  will be considered later, which is defined as
$x_i \Lambda_{ij} x_j > 0$ for any nonzero
real $x_i$. The positive lattice will insure that $\theta$ function
in the equation \aaj is well defined.
The dual lattice $e^{\ast}_i$ is defined as
$$e_i^{\ast} \cdot e_j =\delta_{ij} \, , \eqn\aal$$
and they have inner products as $e_i^{\ast} \cdot e^{\ast}_j =
\Lambda^{-1}_{i,j}$. \par
It can be verified that
$$\eqalign{& \theta {a\brack b} (z+e_i|e, \tau )=e^{2\pi ia\cdot e_i}
\theta {a\brack b} (z|e, \tau )\, , \cr & \theta {a\brack b}
(z+\tau e_i|e, \tau )=\exp {[-\pi i \tau e_i^2 -2\pi i e_i \cdot (z+b)]}
\theta {a\brack b} (z|e, \tau ) \, , \cr
& \theta {a\brack b} (z+e^{\ast}_i|e, \tau )=e^{2\pi ia\cdot e^{\ast}_i}
\theta {a\brack b} (z|e, \tau )\, , \cr & \theta {a\brack b}
(z+\tau e^{\ast}_i|e, \tau )=\exp {[-\pi i \tau {(e_i^{\ast})}^2
-2\pi i e^{\ast}_i \cdot (z+b)]}
\theta {a+e^{\ast}_i\brack b} (z|e, \tau ) \, , \cr}\eqn\aak$$
and
$$\theta {a+e_i\brack b+e^{\ast}_j} (z|e, \tau )=\exp (2\pi i a \cdot
e_j^{\ast}) \theta {a\brack b} (z|e, \tau ) \, . \eqn\aar$$
Moreover in {\it 1-dimension} lattice with $e_1 \cdot e_1 =1$, we define
$$\theta_3(z|\tau)=\theta {{1\over 2}\brack {1\over 2}}
(z|\tau )\, , \eqn\aal$$
which is an odd function of $z$. It can be verified that
$$\eqalign{& \theta_3(z+1|\tau)=e^{\pi i}
\theta_3(z| \tau )\, , \cr & \theta_3
(z+\tau | \tau )=\exp {[-\pi i \tau  -2\pi i \cdot (z+{1\over 2})]}
\theta_3(z|\tau ) \, . \cr} \eqn\aam$$
The Laughlin-Jastrow wave functions on the torus at the filling $1\over m$
with $m$ being an odd positive integer can be written as
$$\eqalign{&\Psi (z_i)=\exp (-{\pi \Phi \sum_i y^2_i \over
\tau_2})F(z_i)\, ,\cr & F(z_i)=\theta {a\brack b} (\sum_i z_ie|e,\tau)
\prod_{i<j} {[\theta_3(z_i-z_j|\tau)]}^m \, , \cr} \eqn\aan$$
where $\theta$  function is on {\it 1-dimension} lattice, $e^2=m$ , $i=1, 2
\ldots ,  N$ with $N$ being the number of the electrons
and $a=a_1e, b=b_1e$. Thus
$$\eqalign{&F(z_i+1)={(-1)}^{N-1}e^{2\pi a_1 m}
F(z_i)\, , \cr & F(z_i+\tau)=\exp (-\pi (N-1)-2\pi i mb_1)
\exp [-i\pi mN(2z_i+\tau)] F(z_i)\, .\cr} \eqn\aao$$
Comparing to the equation \aah , we get
$$\Phi=mN, \phi_1=\pi  (\Phi +1)+2\pi n_1 +2\pi a_1m ,
\phi_2=\pi (\Phi +1)+2\pi n_2 -2\pi b_1m \, . \eqn\aap$$
Thus the solutions of \aap will give $m$ ground wave functions,
which  can be shown by using the equation \aar .
Explicitly the solutions are
$$a_1={\phi_1 \over 2\pi m}+{\Phi +1\over 2m}+
{i\over m},  b_1=-{\phi_2 \over 2\pi m}+{\Phi+1\over 2m} \, \, ,
i=0,1,\ldots , m-1  \, , \eqn\aas$$
which will give  $m$ orthogonal Laughlin-Jastrow wave functions.
So there is $m$-fold center-mass degeneracy [\HR]
(its possible physical relevance has been studied in [\HRN]).
\par
We give a remark here.
Different ground states are connected by gauge transformation of the magnetic
field [\ROBERTO,\ILN].
The degeneracy will disappear and one unique ground state appears
by fixing the gauge of the magnetic field.
Let us use the above Laughlin
wave functions as an example. Now we take the magnetic potential as
$A_1=(-B y+{2\pi c_1\over \tau_2})$\^x, $ A_2={2\pi c_2\over \tau_2}$\^y
with $c_1, c_2$ being constant and
$c=ic_1-c_2$. The magnetic field is independent on  constants $c_i$.
That $c_i$ can be any constant numbers  and the choice of $c_i$ is
the choice of the gauge of the magnetic potential. $c$ and $c+m+n$
are gauge equivalent as they are connected by
the {\it large gauge} transformations on the torus which are generated by
$U_1=\exp ({-2\pi iy\over \tau_2})$ and
$U_2=\exp [{\pi (\tau {\bar z}-{\bar \tau} z )\over \tau_2}]$. If we take
$A_1=(-B y+{2\pi c_1\over \tau_2})$\^x, $ A_2={2\pi c_2\over \tau_2}$\^y,
the wave functions   will be
$$\eqalign{&\Psi (z_i)=\exp
({\pi \over 2\tau_2}[ce^{\ast}+\sum_i ({\bar z}_i-z_i)e]^2)  \cr &
\phantom{\Psi (z_i)=} \cdot
\exp (-{\pi \Phi \sum_i y^2_i \over \tau_2}-{\pi \over 2\tau_2}
[\sum_i ({\bar z}_i-z_i)e]^2) F(z_i)\, ,
\cr & F(z_i)=\theta {a\brack b} (\sum_i z_ie -ce^{\ast}|e,\tau)
\prod_{i<j} {[\theta_3(z_i-z_j|\tau)]}^m \, , \cr} $$
where $e^{\ast}={1\over e}$ and $a, b$
are remain unchanged and still given by \aas.
Then we find that the above wave functions will  be invariant up to a phase
under the related {\it large  gauge} transformation of $U_1$ and
will change from one to another
under  the related {\it large  gauge} transformation of $U_2$.
So we can say that
the degeneracy will disappear  by fixing the gauge of the magnetic field.
The definition of the large gauge transformation on the wave functions can
be found in [\ILN] or formula $(4.72)$ in [\ROBERTO], see also [\PRE]
(but pay attention on the different notation).
\par
In [\ROBERTO,\ILN],
they also show that the phases
$\phi_1 \, , \phi_2$  need to be fixed if requiring that
the wave function under the modular transformations will be
transformed covariantly.
We will come back to this point at the end of the paper.
\par
The hierarchical FQHE, similar to the one considered by N. Read [\REA] on the
plane,  can be characterized by the symmetric matrix
$\Lambda$ with $\Lambda_{1,1}$ being an odd positive integer and other diagonal
elements being even integers. In particular  we consider an integer
lattice defined by matrix
$$\Lambda =\pmatrix{p_1&-1&0&\ldots&0&0\cr
-1&p_2&-1&0&\ldots&0\cr
0&-1&p_3&-1&0&\ldots\cr
\vdots&\vdots&\ddots&\ddots&\vdots&\vdots\cr
\vdots&\vdots&\ddots&\ddots&\vdots&\vdots\cr
0&\ldots&0&-1&p_{l-1}&-1\cr
0&0&\ldots&0&-1&p_l\cr} \, , \eqn\aat$$
where $p_1$ is a positive integer, $p_i, i=2, 3, \ldots, l$ are even
integers. $\Lambda$ describes a {\it l-level} hierarchical state.
The coordinates of the particles are expressed by  $z^s_i$, where $z^s_i$
is the coordinate of the $i^{th}$ particle in level $s$, for example,
$z^1_i=z_i$ is the coordinate of the $i^{th}$ electron.
The wave functions  on the torus are supposed to be
$$\eqalign{F(z_i)&=\int \prod_{s=2, i}^{s=l}dz^s_i d{\bar z}^s_i
\theta {a\brack b} (\sum_{s=1}^l\sum_i z^s_ie_s|e,\tau) \cr &\phantom{=}
\cdot \prod_{i, j, s(1)\leq s(2)} {[\theta_3(z^{s(1)}_i-
z^{s(2)}_j|\tau)]}^{e_{s(1)}\cdot e_{s(2)}} \, . \cr} \eqn\aau$$
The  variables $z^s_i, s=2, 3, \ldots , l$
are integrated in the region of the torus and $i, j, s(1)\leq s(2)$
means that if $s(1)=s(2)$, then we take $i<j$.
The explicit form of $e_i$ is not important because the final results
only depend on the inner products among $e_i$. In order that
the function integrated in \aau is
mathematical well defined on the torus, it is  required that the
integrated function is periodic with coordinates
$z^s_i, s=2, 3, \ldots , l$ around the non-contractible loops
of the torus.
\par
In order that the $\theta$ function in \aau be well defined,
the matrix $\Lambda$ must be positive definite. So
$p_i, i=2, 3, \ldots, l$ must be positive even integers.
\par
By applying  the condition \aah , we get
one of those requirements,
$$\sum_j \Lambda_{ij}N_j=\cases {\Phi, &if $i=1$; \cr 0, &otherwise, \cr}
\eqn\aav$$
where $N_j$ is the number of the particles in level $j$.
The filling factor equals to $\nu={N_1 \over \Phi}=e^{\ast}_1\cdot e^{\ast}_1
=\Lambda^{-1}_{1,1}$,   where $N_1$ is the number of the electrons,
and  we find from \aav
$$\nu ={1\over \displaystyle p_1-
{\strut 1\over \displaystyle p_2-
{\strut 1\over \displaystyle \cdots -
{\strut 1\over \displaystyle p_l}}}}  \, .\eqn\filling$$
It is also possible to construct other hierarchy states [\BW],
and we  will discuss it
on the torus elsewhere [\PRE]. Here we just remark  that
the filling factor for this case  turns out to be
$$\nu ={1\over \displaystyle p_1+
{\strut 1\over \displaystyle p_2+
{\strut 1\over \displaystyle \cdots +
{\strut 1\over \displaystyle p_l}}}}  \, ,\eqn\fillingn$$
where $p_1$ is an odd positive integer and $p_i$ with $i\not=1$ is a
positive even integer. Actually, the filling factors in the equations
\filling , the filling of charge conjugate state of the state
with filling given by \filling (about conjugate state, see
[\FRACT] and [\PRE]),
and \fillingn cover almost all kinds of  fillings
realized in the experiments with
the filling having an odd denominator [\XIE].
The  filling  observed in experiment quoted in [\XIE], which does not
belong to the above category, is $3\over 11$.
The mathematical requirement for the well defined wave functions
restricts the possible types of the filling factors of those kinds
of hierarchical wave functions
and this fact maybe can explain why   \fillingn ,
the filling factors in
the equations \filling and the filling
factors of its conjugate
cover almost all fillings observed in the experiments.
\par
Let us continue the discussion of the wave functions \aau .
Another requirement for the periodicity of the function \aau  is
$$\eqalign{&e_s \cdot b=integer \, ,
e_s \cdot a=integer \, , s=2,3, \ldots , l \, ,\cr
&\phi_1=\pi (\Phi+1) +2\pi e_1 \cdot a +2\pi \cdot integer \, ,\cr
&\phi_2=\pi (\Phi+1)-2\pi e_1 \cdot b +2\pi \cdot integer  \, .\cr} \eqn\aay$$
By writing  $a, b$ as $a=a^s e^{\ast}_s \, , b=b^s e^{\ast}_s $
in the bases of the lattice $\Lambda^{\ast}$
(the inverse lattice of $\Lambda$),    \aay
become
$$\eqalign{&a^1={\phi_1 \over 2\pi}+{\Phi +1\over 2}+n_1 \, ,\cr &
b^1=-{\phi_2 \over 2\pi}+{\Phi +1\over 2}+n_2  \, ,\cr &
a^s, b^s =integer, \, \, s=2, 3, \ldots , l \, . \cr} \eqn\aaz$$
The difference between two solutions of $a, b$ lies on $n^s e^{\ast}_s$
with $n^s$ being integers.  The two  solutions of $a$ will give two
orthogonal wave functions, if the difference of two $a$ does not lie on
$k^s e_s$ with $k^s$ being integers, the two solutions of $b$ with the same $a$
will give the same wave function, as it can be shown by using \aar.  So
all wave functions of Laughlin-Jastrow type are described by the solutions
$$\eqalign{&a=[{\phi_1 \over 2\pi}+{\Phi +1\over 2}]e^{\ast}_1
+\sum_s n^se^{\ast}_s  \, ,\cr &
b=[-{\phi_2 \over 2\pi}+{\Phi +1\over 2}]e^{\ast}_1  \, , \cr &
n^se^{\ast}_s \subset {\Lambda^{\ast}\over \Lambda} \, . \cr} \eqn\bbb$$
${\Lambda^{\ast}\over \Lambda}$ means the space $n^se^{\ast}_s$ with $n^s$
being integers and identifying $a(1)=n^s(1)e^{\ast}_s$ with
$a(2)=n^s(2)e^{\ast}_s$ if $a(1)-a(2)=k^se_s$ with $k^s$ being integers.
There are  $\det \Lambda$ solutions to \bbb.
So the degeneracy of the
ground states is $\det \Lambda$,  which actually is the denominator
of the filling.
It  has been predicted in [\WQ] by using
Ginzburg-Landau-Chern-Simons theory of FQHE and in [\GREAD]
by using Conformal Field theory. However
they [\WQ,\GREAD] did not give explicit hierarchical
wave functions on the torus.
\par
To consider the excitations of the states given by \aau, we take the
simplest case as an example,
one quasiparticle at $z$. Thus the wave functions \aau become
$$\eqalign{F(z_i)&=\int \prod_{s=2, i}^{s=l}dz^s_i d{\bar z}^s_i
\theta {a\brack b} (\sum_{s=1}^l\sum_i z^s_ie_s
+z{\bf q}|e,\tau) \cr &\phantom{=} \times
\prod_{i, j, s(1)\leq s(2)} {[\theta_3(z^{s(1)}_i-
z^{s(2)}_j|\tau)]}^{e_{s(1)}\cdot e_{s(2)}}\cr &\phantom{=} \times
\prod_{i, s}{[\theta_3(z-z^s_i|\tau)]}^{{\bf q} \cdot e_s}
\, , \cr} \eqn\bbc$$
where ${\bf q}=q^se^{\ast}_s$ with $q^s$ being integers. ${\bf q}$ will
characterize the statistics and charge of the quasiparticle.
The equation \aav now changes to
$$\sum_j \Lambda_{ij}N_j+e_i \cdot {\bf q}
=\cases {\Phi, &if $i=1$; \cr 0, &otherwise. \cr} \eqn\bbd$$
The equation  \aay, \aaz, \bbb remain unchanged.
It can be shown
$$\eqalign{F(z+1|a,b)&= F(z|a,b) \exp (2\pi i a\cdot {\bf q}+\pi i \Phi
e^{\ast}_1 \cdot {\bf q}-\pi i {\bf q}^2) \, , \cr
F(z+\tau {\bf q}|a,b)&=F(z|a+{\bf q},b)\exp (-\pi i
{\bf q}\cdot e^{\ast}_1\Phi (2z+\tau) \cr &-2\pi i{\bf q}\cdot b
+\pi i{\bf q}^2-\pi i {\bf q}\cdot e^{\ast}_1 \Phi)\, . \cr} \eqn\bbe$$
The charge of the quasiparticle is $e^{\ast}_1 \cdot {\bf q}$.
and the statistics parameter of the quasiparticle $e^{i\theta}$
obtained by interchanging two identical excitations, is given by
$\theta =\pi {\bf q}^2$.
\par
The normalized wave functions of \bbc should contain a normalization
constant which depends on the coordinate $z$.
The normalized wave functions are
$$\Psi_n(z_i, z)=\Psi(z_i, z)\exp {(-{\Phi e^{\ast}_1 \cdot {\bf q} y^2
\over \tau_2})} \, . \eqn\bbf$$
The similar discussion about the normalization in
the presence of the quasiparticles  on the
plane and the sphere was discussed in [\BW,\LIN,\LINN]. This will be important
for obtaining the hierarchical wave functions studied in [\BW],
which has also rather  clear physical meaning and  will be discussed in [\PRE].
\par
It will not be surprising that the wave functions of the FQHE relate to
the wave functions of Chern-Simons theory after so many works
[\GP,\ZHK,\RE,\BW,\REA,\GREAD,\ROBERTO].
The functions, which are integrated by the coordinates of
the quasiparticles to get the wave functions of the FQHE, are
$$\eqalign{\Psi_b(z^s_i)&=\exp (-{\pi \Phi \sum_i y^2_i \over\tau_2})
\theta {a\brack b} (\sum_{s=1}^l\sum_i z^s_ie_s|e,\tau)  \cr
&\phantom{=}\cdot \prod_{i, j, s(1)\leq s(2)} {[\theta_3(z^{s(1)}_i-
z^{s(2)}_j|\tau)]}^{e_{s(1)}\cdot e_{s(2)}} \, . \cr} \eqn\bbh$$
They can be written as
$$\eqalign{\Psi_b(z^s_i)&=\exp (-{\pi  {(\sum_{i,s} y^s_i e_s)}^2
\over\tau_2}) \theta {a\brack b} (\sum_{s=1}^l\sum_i z^s_ie_s|e,\tau)
\cr &\phantom{=} \cdot
\prod_{i, j, s(1)\leq s(2)} \exp ({1\over 2}P(z^{s(1)}_i-
z^{s(2)}_j|\tau)e_{s(1)}\cdot e_{s(2)})\cdot unitary\, phase
\, . \cr} \eqn\bbi$$
If take the magnetic potential as $A_1=(-B y+{2\pi c_1\over \tau_2})$\^x,
$ A_2={2\pi c_2\over \tau_2}$\^y, then \bbi becomes
$$\eqalign{\Psi_b(z^s_i)&=\exp ({\pi \over 2\tau_2}
[ce^{\ast}_1+\sum_i ({\bar z}^s_i-z^s_i)e_s]^2
+{\pi  {(\sum_{i,s} y^s_i e_s)}^2  \over\tau_2})
\cr &\phantom{=} \cdot
\prod_{i, j, s(1)\leq s(2)} \exp ({1\over 2}P(z^{s(1)}_i-
z^{s(2)}_j|\tau)e_{s(1)}\cdot e_{s(2)}) \cr &\phantom{=}
\cdot \theta {a\brack b} (\sum_{s=1}^l\sum_i z^s_ie_s-ce^{\ast}_1|e,\tau)
\cdot unitary\, phase  \, , \cr} \eqn\bbii$$
which also are invariant up to a phase under the large gauge
transformation $U_1$ and change from one state to another under the large gauge
transformation $U_2$ (for Laughlin state at simple filling,
see the discussion before). The unitary phase in \bbi and \bbii depends on the
coordinates $z^s_i$. $P(z)$ is the scalar propagator, satisfying
$${1\over 4}\nabla^2 P(z)={\bar \partial}\partial P(z)=\pi (\delta^2(z)
-{1\over \tau_2}) \, ,$$
and is given by
$$P(z)=\ln \left| \theta {{1\over 2} \brack {1\over 2}}(z|\tau) \over
\theta^{\prime} {{1\over 2} \brack {1\over 2}}(0|\tau)\right|^2+
{\pi \over 2\tau_2}{(z-{\bar z})}^2 \, . $$
However it can be shown that wave functions \bbii
are the  wave functions of the generalized Abelian Chern-Simons
theory (up to a unitary phase which actually
relates to singular gauge transformation)
on the multidimensional $R^l/\Lambda$ compact Abelian gauge group,
where $\Lambda$ is the integer lattice with bases $e_i$ defined by
$e_i \cdot e_j =\Lambda$.  The large component of
Chern-Simons potential here is equal to
$c e^{\ast}_1$  [\MACO].
The wave functions of Abelian Chern-Simons theory
on the torus have been studied detailed in [\ILN] and
the discussion of the wave functions of the  generalized
Abelian Chern-Simons theory on the torus can be found in [\MACO].
The lagrangian of the generalized Abelian Chern-Simons
theory which we are interested in,  is given by
$$L=\int dv [-{1\over 4\pi}\epsilon^{\mu \nu \lambda}
 A_{\mu} \cdot \partial_{\nu}
A_{\lambda} + A_0 \cdot (\sum_{s,i}e_s \delta^2(z-z^s_i) -
{1\over 2\pi} \cdot Be^{\ast}_1)]  \, ,\eqn\dual$$
where $A_{\mu}=A_{s,\mu}e_s$ and with the condition
$$\int dv (\sum_{s,i}e_s \delta^2(z-z^s_i) -
{1\over 2\pi} \cdot Be^{\ast}_1) =0 \, . \eqn\conii$$
The equation \conii can be written as
$$\sum_s N_s e_s -\Phi e^{\ast}_1 =0 \, , \eqn\coinn$$
which is the same as  the equation \aav.
The lagrangian \dual will give the right statistics
for the electrons and the quasiparticles.  To introduce an excitation,
we simply add a term $A_0 \cdot {\bf q} \delta^2 (z-z_q)$, where
${\bf q}=q^se^{\ast}_s$  with $q^s$ being integers.
Then we have
$$\sum_s N_s e_s + {\bf q}-\Phi e^{\ast}_1 =0 \, , \eqn\cccc$$
which is the same as  the equation \bbd .
\par
This kind  form of the generalized Abelian Chern-Simons action appears in
dual form Ginzburg-Landau-Chern-Simons theory of FQHE [\BW].
The generalized Abelian Chern-Simons \dual
characterizes the essential properties  of the
hierarchical FQHE, from which  we can obtain the degeneracy of
the ground state and the  excitation spectrums, etc..
\par
To discuss the modular property, let us introduce
$\theta_1 =[{\phi_1 \over 2\pi}+{\Phi +1\over 2}]e^{\ast}_1,
\theta_2 =[-{\phi_2 \over 2\pi}+{\Phi +1\over 2}]e^{\ast}_1$
(this notation is the same as [\MACO]).
If the full modular invariance (covariance) is required,
then we need to choose
$\theta_1=\theta_2={e^{\ast}_1\over 2}$ [\MACO,\ILN].
We may also remark that if
$\theta_1, \theta_2$  are  zero,
the wave functions of the FQHE will be transformed covariantly only under
the {\bf subgroup} of the  modular transformation
$ \tau \to {\alpha \tau +\beta \over \gamma \tau +\delta}$
which is generated by $\tau \to \tau +2 \, ,and \, \tau \to -{1\over \tau}$
(the corresponding coordinate transformation under the modular transformation
is $z \to {z\over \gamma \tau +\delta}$)\foot{This also has been
observed (for the case of the filling $\nu ={1\over m}$) in [\NAP].}.
Above results can be arrived by using
the method  developed, for example  in [\MACO,\ILN].
\par
In summary, we have obtained a kind of
the hierarchical Laughlin-Jastrow wave functions
on the torus. The degeneracy of the ground states is obtained.
The relation of the wave functions with the one in Chern-Simons theory
is revealed. We hope that the wave functions of FQHE on the surface with
arbitrary genus  can be obtained by using the corresponding
results in Chern-Simons  theory.
\par
The author is grateful to Professor R. Iengo  for many stimulating discussions
and constant encouragements, and to Dr. Marco A.C. Kneipp for explaining his
work [\MACO].

\refout
\end
\bye